\documentclass{ws-ijqi}

\usepackage{graphics}

\begin{document}

\markboth{M. Lasota et al.}
{Quantum key distribution with realistic heralded single photon sources}

\title{QUANTUM KEY DISTRIBUTION WITH REALISTIC HERALDED SINGLE PHOTON SOURCES}

\author{Miko\l aj Lasota}

\address{Faculty of Physics, Astronomy and Applied Informatics, Nicolaus Copernicus University\\ul. Grudziadzka~5, 87-100~Toru\'{n}, Poland\\miklas@fizyka.umk.pl}

\author{Rafa{\l} Demkowicz-Dobrza\'{n}ski}

\address{Institute of Theoretical Physics, Faculty of Physics, University of Warsaw\\ul. Ho\.{z}a~69, 00-681~Warsaw, Poland\\Rafal.Demkowicz-Dobrzanski@fuw.edu.pl}

\author{Konrad Banaszek}

\address{Institute of Theoretical Physics, Faculty of Physics, University of Warsaw\\ul. Ho\.{z}a~69, 00-681~Warsaw, Poland\\Konrad.Banaszek@fuw.edu.pl}

\maketitle

\begin{abstract}
We analyze theoretically performance of four-state quantum key distribution protocols
implemented with a realistic heralded single-photon source. The analysis assumes a noisy model for the detector heralding generation of individual photons via spontaneous parametric down-conversion, including dark counts and imperfect photon number resolution. We identify characteristics
of the heralding detector that defines the attainable cryptographic key rate and the maximum secure distance. Approximate analytical formulas are applied to multiplexed detection and compared with results of numerical calculations.
\end{abstract}

\keywords{quantum cryptography, heralded single photon sources, multiplexing detectors}

\section{Introduction}

The security of quantum key distribution (QKD) is guaranteed by the fundamental impossibility to discriminate with certainty non-orthogonal quantum states of an individual physical system.\cite{BB84,GisiRiboRMP02} Single photons are an obvious choice for such a system owing to easy preparation and detection of non-orthogonal states, combined with stable and robust long-haul transmission in optical fibers as well as in free space.\cite{EisaFanRSI11} This makes single photon sources an essential component of many practical QKD setups.\cite{ScarBechRMP09}

Realistic single photon sources suffer from two imperfections. The first one is occasional generation of two or more photons within a given temporal bin. This opens up a possibility of a successful eavesdropping attack based on photon number splitting, when an eavesdropper is able to capture and measure the state of one of the transmitted photons without revealing her presence. The second imperfection is the production of empty vacuum bins, when no photon is actually present. This lowers the key generation rate and, more importantly, increases the relative contribution of dark counts produced by detectors used in the receiver setup. If the security analysis conservatively attributes errors introduced by dark counts to the action of an eavesdropper, this ultimately limits the maximum distance of secure communication.

The simplest approximate realisation of a single photon state is a weak coherent pulse (WCP) derived from a heavily attenuated pulsed laser beam. The Poissonian statistics for the number of photons carried by such a pulse leads to a trade-off between the importance of multiphoton events and vacuum events. This defines the optimal pulse intensity to be used in a QKD protocol.\cite{BrasLutkPRL00} A more advanced approach is to use a heralded single-photon source based on the process of spontaneous parametric down-conversion (SPDC), in which pump photons spontaneously decay into photon pairs.\cite{EisaFanRSI11,BridDegiAPL12} Ideally, detecting one, idler photon from a pair unambiguously heralds the availability of the second, signal photon that can be subsequently used for a cryptographic protocol. This removes the vacuum component from the generated state.\cite{BrasLutkPRL00} Further, using a photon number resolving detector for heralding would eliminate the multiphoton component in the down-converted light beam. Consequently, the down-conversion process could be operated at higher pump powers, increasing the effective generation rate for single photons.

The main purpose of this paper is to analyze theoretically the performance of realistic heralded single photon sources based on SPDC and photon number resolving detectors in the standard BB84 protocol. In practice, the attractive features of heralded sources can be compromised by a number of imperfections. Typical down-conversion sources do not guarantee perfect photon number correlations between signal and idler photons: usually, there is a trade-off between collection efficiencies for the signal and the idler arms, especially when bulk nonlinear media are employed.\cite{KurtOberPRA01,LjunTengPRA05,KoleWasiPRA09} Further, non-unit detection efficiency of the heralding detector undermines its photon number resolving capability, while dark counts in the heralding arm result in vacuum events in the signal arm. Starting from a general model for a realistic photon number resolving detector, we will identify its characteristics that is relevant for QKD applications. This generalizes earlier results obtained for a heralded source with a binary on/off detector.\cite{BrasLutkPRL00} The introduced formalism allows us also to discuss selected security aspects of other qubit-based four-state protocols, such as SARG04\cite{SARG04} which offers partial protection against multiphoton events.

In our analysis we leave aside protocols exploiting the decoy-state method\cite{HwangPRL03,WangPRL05,LoMaChenPRL05,MaQiPRA05} which estimates the contribution from multiphoton events by preparing light pulses with variable photon statistics and measuring corresponding detection probabilities. Such a strategy can enhance substantially the scaling of the key rate with the transmission of the optical channel from quadratic to a linear one. SPDC can be used in this approach as a universal source to generate approximate single photons along with required decoy states,\cite{MaurSilbPRA07,AdaYamaPRL07,MaLoNJP08,CurtyMaPRA10} conditioned upon the measurement outcome on the heralding detector. One can also consider utilizing heralded single photons generated by SPDC as signal pulses,\cite{HoriKobaPRA06} although optimization over the pump power would yield unrealistically high requirements for the efficiency of the down-conversion process. A way out would be to arrange an array of weakly pumped down-conversion sources\cite{WangChenPRL08} followed by a multiport switch picking up a photon from a source flagged by the respective heralding detector. However, the complexity of such a setup would lead in practice to additional losses that might easily compromise its performance. Decoy-state techniques require a careful analysis of a much broader range of eavesdropping strategies in multiple degrees of freedom,\cite{HelwMauePRA09} and their statistical aspects are significantly more complex\cite{WangPRL05,MaQiPRA05,MaLoNJP08,TanCaiIJQI11} compared to the effects of finite-length data in standard single-photon protocols.\cite{FiniteKeyLength,RennerIJQI08} This motivated us to reexamine the performance of heralded single photons in quantum key distribution under the conservative assumptions of non-decoy schemes.

This paper is organized as follows. In Sec.~\ref{Sec:KeyRate} we introduce basic expressions that characterize the performance of the QKD protocol with a heralded single-photon source. The short-distance limit, when dark counts can be neglected, is discussed analytically in Sec.~\ref{Sec:HighTransmission}. In Sec.~\ref{Sec:MaximumDistance} we give approximate expressions for the maximum secure distance. These results are compared in Sec.~\ref{Sec:Numerical} with numerical results and the advantages of multiplexed heralding are discussed. Finally, Sec.~\ref{Sec:Conclusions} concludes the paper.

\section{Key rate}
\label{Sec:KeyRate}

We will consider here the standard physical layer of the BB84 protocol: Alice prepares one of four qubit states that form two mutually unbiased bases, while Bob switches between projective measurements in one of these two bases and directs two orthogonal states to separate detectors. For simplicity we assume that errors are the same in both bases,\cite{GLLP} which is a reasonable model for phase and polarization encodings. The starting point of our discussion will be an expression for the key rate $K$ as a function of the quantum bit error rate (QBER) $Q$ determined by Alice and Bob from the statistics of events, and the fraction $y$ of detection events generated by pulses that genuinely contained a single photon. We assume that Eve performs a photon number splitting attack, capturing one photon and letting the second one travel without any disturbance. We will approximate her information gain for multiphoton pulses by its value for two-photon cases $I_{AE}^{(2)}$, taken as independent of $Q$.
We also assume that the observed QBER stems entirely from an attack on single-photon pulses, and that Eve prevents any detection events to occur at Bob's side when Alice sends a vacuum pulse. These assumptions lead to the following expression for the secure key rate
\begin{equation}
\label{Eq:KeyRate}
K = p_{\rm{exp}} p_{\rm{sift}} [ I_{AB}(Q)- y I_{AE}^{(1)} (Q/y) - (1-y) I_{AE}^{(2)}],
\end{equation}
where $p_{\rm{exp}}$ is the total expected probability of a detection event, $p_{\rm{sift}}$ is the fraction of all the events contributing to the raw key, and $I_{AE}^{(1)}(Q/y)$ is Eve's information gain for single-photon pulses, calculated for the QBER rescaled by $y$
to account for the multiphoton fraction that is assumed to introduce no errors. The quality of the raw key is characterized by the mutual information $I_{AB}(Q)$  between Alice and Bob, equal to
\begin{equation}
I_{AB}(Q) = 1 - H(Q),
\end{equation}
where $H(x) = - x\log_2 x - (1-x) \log_2 (1-x)$ is the binary entropy. For the standard BB84 protocol we have $p_{\rm{sift}}=1/2$, as bases chosen by Alice and Bob are compatible in 50\% of the cases. Eve's collective attack on single-photon pulses yields mutual information equal to $I_{AE}^{(1)}(Q) = H(Q)$,\cite{GisiRiboRMP02} while the photon number splitting attack makes two-photon pulses completely insecure and consequently $I_{AE}^{(2)} =1$.

The formula assumed in Eq.~(\ref{Eq:KeyRate}) can be applied also to certain eavesdropping strategies for other protocols. As an example we will consider here the SARG04 protocol. In this case we will take $p_{\rm{sift}} \approx 1/4$, neglecting a minor change in this value for a non-zero QBER. Eve's collective attack on single-photon pulses results in her information gain described by an expression\cite{BranGisiPRA05}
\begin{eqnarray}
I_{AE}^{(1)} (Q) & = & (1-Q) \log_2 (1-Q) - (1-2Q) \log_2(1-2Q) + Q (1-\log_2 Q).
\end{eqnarray}
Further, if Eve applies the photon number splitting attack to two-photon pulses and attempts to infer the key bits only from retained photons, her task is more difficult than in the BB84 protocol since the information revealed publicly by the legitimate parties does not identify unambiguously the measurement basis. In fact, having learned the public information, Eve is left with the problem of discriminating between two equiprobable non-orthogonal states with the absolute value of their scalar product equal to $1/\sqrt{2}$. Consequently, her maximal information gain is bounded by the Holevo quantity\cite{Holevo} calculated for this ensemble, which yields the value
\begin{equation}
I_{AE}^{(2)} = H \bigl( (2+\sqrt{2})/4 \bigr) \approx 0.6009.
\end{equation}
We do not exploit here the possibility of extracting the key from multiphoton events.\cite{TamaLoPRA06}

In the following, key rates calculated using the two sets of formulas for the BB84 and SARG04 protocols will be denoted respectively as $K_{\rm{BB84}}$ and $K_{\rm{SARG04}}$. It should be noted that the eavesdropping model considered here makes sense as long as Eve's information gain from multiphoton pulses exceeds that from single photon events, i.e.\ $I_{AE}^{(1)} (Q/y) \le I_{AE}^{(2)}$. If this condition does not hold, the photon number splitting attack is not effective and it should be replaced by a more general multiphoton attack. For the BB84 protocol, the condition $I_{AE}^{(1)} (Q/y) \le I_{AE}^{(2)}$ is satisfied automatically, as multiphoton events are completely insecure. In the case of the eavesdropping strategy for the SARG04 protocol described above one needs to restrict applicability to sufficiently low ratios $Q/y$. This issue will be highlighted in Sec.~\ref{Sec:MaximumDistance}. The consistency of the obtained results with the assumptions made in the security analysis should be checked numerically when applying results derived in this paper to a specific experimental scenario.

In the next step, we need to relate the parameters $p_{\rm{exp}}$, $Q$, and $y$ to the properties of the actual physical setup. We will consider a heralded single photon source based on a multi-mode SPDC process, in which the statistics of the number of generated pairs is well approximated by the Poissonian distribution.\cite{HelwMauePRA09} Consequently, the probabilities $p_n$ of generating $n=0$ and $n=1$ pairs can be written respectively as
$p_0 = e^{-\lambda}$ and $p_1 = \lambda e^{-\lambda}$, where $\lambda \ll 1$ is defined by the pump power and the strength of the nonlinear interaction. The probability of producing two or more pairs is given by $p_{2} = 1-p_0-p_1 = 1 - (1+\lambda) e^{-\lambda}$. In the perturbative regime of weak pumping the dominant contribution to multiple events comes from generation of double photon pairs, and our security analysis will assume that all multiple events are of this form.

One of the down-converted photons, used as a herald, is measured by a detector on Alice's side. Only events in which Alice's detector signalled the presence of exactly one photon are retained by Alice and Bob. Let us denote by $q_i$, $i=0,1,2$, the conditional probability that the heralding detector generated this outcome, provided that it received $i$ photons. For an ideal photon number resolving detector we have $q_1=1$ and $q_0 = q_2 =0$. Deviations from this perfect characteristics will have a critical impact on the performance of the source. The results for weak coherent pulses are recovered for the no-heralding case $q_0=q_1=q_2=1$.

Two other relevant parameters are the transmission of the optical channel from Alice to Bob multiplied by Bob's detection efficiency, denoted here by a single parameter $T$, and the probability of a dark count on each of Bob's two detectors $d_B \ll 1$. The expected overall probability of a detection event is given by an expression:
\begin{equation}
p_{\rm exp} = Tp_1q_1+ 2 T p_2 q_2 + 2d_B(p_0q_0+p_1q_1 +p_2 q_2) \label{eq:pexpmodif},
\end{equation}
where we ignored double-count events described by terms of the order of $T d_B$, $T^2$ and $d_B^2$. As on average half of dark count events generate errors for both BB84 and SARG04 protocols, the QBER is given by
\begin{equation}
Q  = \frac{d_B}{p_{\rm exp}} (p_0q_0+p_1q_1 +p_2 q_2). \label{Eq:qH}
\end{equation}
This QBER level will be interpreted as a result of Eve's attack on single-photon pulses.
Finally, the fraction of detection events that can be attributed to single photons is
\begin{equation}
y = 1 - \frac{p_2 q_2}{p_{\rm{exp}}}. \label{Eq:yH}
\end{equation}
In the following sections we will discuss performance of QKD in the regime of short distances on one hand and derive an approximate expression for the maximum secure distance on the other hand. These results will be compared with complete numerical calculations based on general formulas introduced above.

\section{Short distance limit}
\label{Sec:HighTransmission}

In order to gain intuition about differences between various realizations of the protocol, let us first consider the short-distance limit when the overall transmission of the optical channel is relatively high so that the number of detection events generated by actual photons substantially exceeds that triggered by dark counts. In this case, simple estimates on the key rates can be obtained by entirely neglecting dark counts on Bob's detectors. Under this assumption, the expression for the key rate simplifies to the form
\begin{equation}
K  =  p_{\exp} p_{\rm{sift}} [ 1 - (1-y) I_{AE}^{(2)}] \nonumber\\ = p_{\rm{sift}} [ T p_1 q_1 - p_2 q_2 ( I^{(2)}_{AE} - 2 T)]  ,
\end{equation}
which can be used for both BB84 and SARG04 protocols by a suitable choice of $p_{\rm{sift}}$ and $I_{AE}^{(2)}$.
In order to optimize the performance of the protocol, we need to maximize the right hand side of the above formula with respect to the pump strength $\lambda$. The maximum is reached at
\begin{equation}
\lambda_{\rm{short}} = \frac{Tq_1}{Tq_1 + ( I^{(2)}_{AE} - 2T ) q_2}.
\end{equation}
In the regime when $Tq_1 \ll ( I^{(2)}_{AE} - 2T ) q_2$ we can approximate $\lambda_{\rm{short}} \approx Tq_1 / [I^{(2)}_{AE} - 2 T]q_2$ and take
$p_1 \approx \lambda_{\rm{short}}$, $p_2 \approx \lambda_{\rm{short}}^2 /2$. This yields the following compact formula for the key rate:
\begin{equation}
K \approx    \frac{q_1^2}{q_2} \frac{ p_{\rm{sift}} T^2}{2 ( I^{(2)}_{AE} - 2 T ) }.
\end{equation}
It will be convenient to use the superscript $^{(C)}$ for quantities specialized to the case of WCPs, while the superscript $^{(H)}$ will denote quantities calculated for a general heralded single photon source. It is seen that properties of the detector enter through the factor $q_1^2/q_2$. Because for WCPs $q_1^2/q_2 =1$, we immediately obtain that
\begin{equation}
\label{Eq:KHSPSapprox}
K^{(H)} = \frac{q_1^2}{q_2} K^{(C)}.
\end{equation}

The above formula provides a simple criterion to characterize the usefulness of the heralded source for short distances. Let us note here that in order to ensure consistency of the above estimates with the perturbative treatment of the photon number distribution, we should assume that $\lambda_{\rm{short}} \ll 1$. This means that the channel transmission is low enough and the detector is sufficiently imperfect to carry out calculations in the perturbative regime, but at the same time the rate of detection events is high enough to neglect the dark counts on Bob's detectors.

\begin{figure}[pt]
\centerline{\resizebox{0.8\textwidth}{!}{\psfig{file=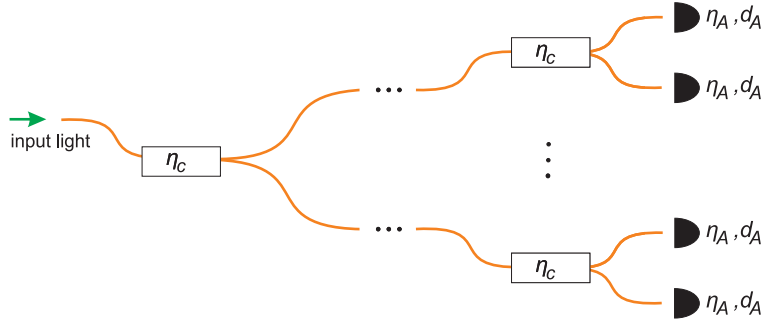}}}
\vspace*{8pt}
\caption{A schematic of a multiplexing detector splitting the input signal using balanced couplers, each with overall transmission $\eta_c$, and directing outputs to individual binary detectors characterized by the efficiency $\eta_A$ and the dark count rate $d_A$.}
\label{Fig:TreeDetector}
\end{figure}

As an exemplary application of Eq.~(\ref{Eq:KHSPSapprox}), let us consider an $N$-stage tree beam splitting scheme shown in Fig.~\ref{Fig:TreeDetector}, which formally is equivalent to time-multiplexed detection.\cite{AchiSilbOL03,FitcJacoPRA03} At each stage, the input light is split into twice as many beams using balanced couplers. The output ports are monitored with binary detectors characterized by the efficiency $\eta_A$ and the dark count probability $d_A$. A straightforward calculation based on classical probability theory yields the following expressions for conditional probabilities $q_i$:
\begin{eqnarray}
q_0 & = & (1-d_A)^{2^N-1}  2^N d_A, \nonumber \\
q_1 & = & (1-d_A)^{2^N-1} [2^N d_A(1-{\eta}_A)+{\eta}_A],\\
q_2 & = & (1-d_A)^{2^N-1} [2^N d_A(1-{\eta}_A)^2+2{\eta}_A(1-{\eta}_A) + 2^{-N}{\eta}_A^2]. \nonumber
\end{eqnarray}
These formulas assume lossless 50/50 couplers used in the multiplexing detector. In order to consider realistic lossy couplers with overall power transmission $\eta_c$ each, we will replace in the above expressions $\eta_A$ by $\eta_A \eta_c^N$.

For high count rates, in the first approximation we can  neglect dark counts also on Alice's detector. Then the ratio $q_1^2/q_2$ is given explicitly by a compact formula
\begin{equation}
\label{Eq:q12q2}
\frac{q_1^2}{q_2} = \frac{1}{2/(\eta_A \eta_c^N) - 2 + 2^{-N}}.
\end{equation}
It is seen that if $\eta_c = 100\%$, increasing the number of stages improves the ratio $q_1^2/q_2$ and the standard expression for a photon number resolving detector is recovered in the limit $N \rightarrow \infty$. This however relies critically on the assumption of perfectly lossless couplers, as otherwise the effective efficiency $\eta_A \eta_c^N$ tends to zero.

The heralded source yields a higher key rate than weak coherent pulses when $q_1^2/q_2 > 1$, which can be written as the following condition on Alice's detector efficiency:
\begin{equation}
\eta_A \eta_c^N > \frac{2}{3- 2^{-N}}.
\end{equation}
For a binary heralding detector, which corresponds to \mbox{$N=0$}, the heralded source cannot bring any advantage over weak coherent pulses in the short distance regime. In the limit of full photon number resolution, when \mbox{$N \rightarrow \infty$}, the efficiency of the heralding detector should exceed \mbox{$\eta_A \eta_c^N > 2/3$} to offer an improvement over weak coherent pulses. As $\eta_A$ includes also the collection efficiency of heralding photons, this poses rather stringent requirements on heralded single photon sources to be useful in the short distance regime.

In the following sections  we will see that heralded sources warrant security over significantly longer distances compared to weak coherent pulses, and that the short-distance approximation presented above works for heralded sources well beyond the maximum secure distance of the WCP scheme. In this case, we need to compare factors $q_1^2/q_2$ calculated for binary and multiplexed heralding. The expression derived in Eq.~(\ref{Eq:q12q2}) suggests that for favourable technical parameters multiplexing can boost the secret key rate.

\section{Maximum distance}
\label{Sec:MaximumDistance}

\begin{figure}[pt]
\centerline{\resizebox{1.0\textwidth}{!}{\psfig{file=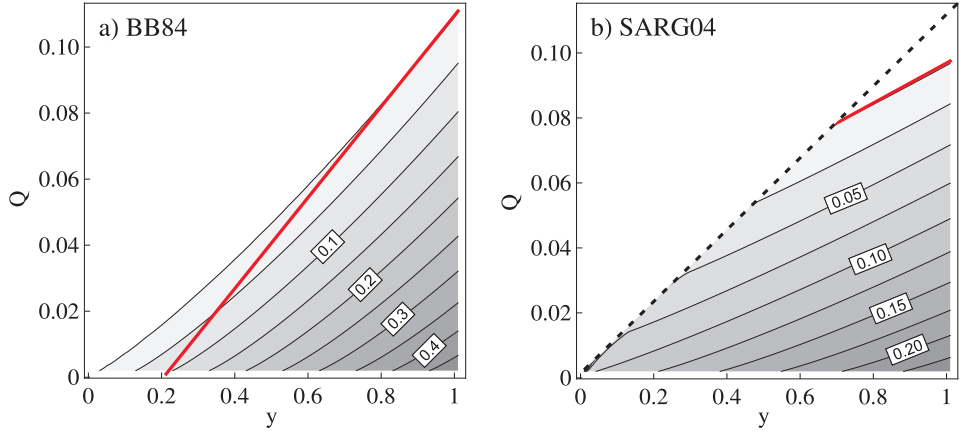}}}
\vspace*{8pt}
\caption{Contour plots of the renormalized key rate for (a) BB84 and (b) SARG04 protocols. In (b), the parameter region not satisfying the assumption $I^{(1)}_{AE}(Q/y) < I^{(2)}_{AE}$ is blanked out and separated with a dashed line. The thick solid lines (red online) depict the linearized necessary condition for the key security given in Eq.~(\protect\ref{Eq:Q<Q(1)}).}
\label{Fig:KeyRateyQ}
\end{figure}

We will now consider the other limiting case and estimate the maximum distance---or equivalently the minimum transmission of the optical channel---over which a secure key can be established. For low transmission, a larger fraction of events is generated by dark counts, which increases the value of QBER. We will focus our attention on the regime when the sender cannot afford to produce too many multiphoton events which would provide Eve with more information than single photons without contributing to the QBER.  In this regime we can linearize the condition for the positivity of the key to the form
\begin{equation}
\label{Eq:Q<Q(1)}
Q < Q^{\rm th}[1 - \xi (1-y)],
\end{equation}
where $Q^{\rm th}$ is the threshold QBER for an ideal single photon source obtained from the equation $I_{AB}(Q^{\rm th}) = I_{AE}^{(1)}(Q^{\rm th})$, and $\xi$ is the proportionality factor that characterizes the effective reduction in this value introduced by the fraction of multiphoton events.
These two parameters depend on the protocol under consideration and their numerical values are $Q^{\rm th}\approx 0.11$ and $\xi\approx 1.25$ for the BB84 protocol and $Q^{\rm th}\approx 0.0968$ and $\xi\approx 0.64$ for the SARG04 protocol. In Fig.~\ref{Fig:KeyRateyQ}, this linearized condition is superimposed on the contour plots of the renormalized key rate $K/p_{\rm{exp}}$ calculated for BB84 and SARG04 protocols. Note that for the SARG04 protocol, the applicability is restricted by the condition $I_{AE}^{(1)} (Q/y) \le I_{AE}^{(2)}$.

Before analyzing the performance of the heralded source, it will be helpful to revisit, using Eq.~(\ref{Eq:Q<Q(1)}) as the starting point, the bounds on minimum transmission for a perfect single-photon source and weak coherent pulses obtained by Brassard {\em et al.} in Ref.~\refcite{BrasLutkPRL00}. In the case of an ideal single photon source, we have $Q = d_B/ (T + 2 d_B)$. Using this expression we can transform Eq.~(\ref{Eq:Q<Q(1)}) to the form $T > T^{(1)}_{\rm{min}}$, where the minimum transmission for a perfect single-photon source ($y=1$), denoted below as $T^{(1)}_{\rm{min}}$, is given by
\begin{equation}
\label{eq:perfect1}
T^{(1)}_{\rm{min}} = d_B\frac{1-2 Q^{\rm th}}{Q^{\rm th}}.
\end{equation}
For WCPs, we need to transform Eq.~(\ref{Eq:Q<Q(1)}) into a lower bound on $T$ and then minimize it over the average photon number, which yields the optimal \mbox{$\lambda_{\rm{opt}}^{(C)} = [2 d_B (1-2 Q^{\rm th})/ Q^{\rm th} \xi]^{1/2}$} and the minimum transmission is equal to
\begin{equation}
T_{\rm min}^{(C)} = \left(\frac{2 d_B \xi  (1-2 Q^{\rm th})}{Q^{\rm th}}\right)^{1/2}. \label{eq:wcp1}
\end{equation}
This expression scales as $d_B^{1/2}$ for the dark count probabilities $d_B \ll 1$ and therefore is much larger than $T^{(1)}_{\rm{min}}$, which depends linearly on $d_B$.

An analogous procedure can be applied to a heralded single photon source. Neglecting terms of the order of $p_2q_2$ in formulas Eqs.~(\ref{eq:pexpmodif}) and (\ref{Eq:qH}) and inserting them along with Eq.~(\ref{Eq:yH}) into Eq.~(\ref{Eq:Q<Q(1)})
gives the following constraint on the transmission $T$:
\begin{equation}
T>  \frac{q_2}{2q_1} \xi \lambda +d_B \frac{1-2Q^{\rm th}}{Q^{\rm th}}+d_B\frac{1-2Q^{\rm th}}{Q^{\rm th}}\frac{q_0}{q_1}\frac{1}{\lambda}.
\label{eq:maininequality}
\end{equation}
The minimum value of the right hand side is achieved for
\begin{equation}
\lambda_{\rm opt}^{(H)} = \left(2d_B \frac{1-2 Q^{\rm th}}{\xi Q^{\rm th}}\frac{q_0}{q_2}\right)^{1/2}. \label{eq:chizero}
\end{equation}
It is easy to see that the explicit expression for the minimum transmission which guarantees secure communication with heralded single photon sources can be formally represented as a combination of $T_{\rm min}^{(1)}$ and $T_{\rm min}^{(C)}$ defined, respectively, in Eqs.~(\ref{eq:perfect1}) and (\ref{eq:wcp1}):
\begin{eqnarray}
T_{\rm min}^{(H)} & = & d_B \frac{1-2Q^{\rm th}}{Q^{\rm th}}+\left( 2 d_B \xi \frac{1-2Q^{\rm th}}{Q^{\rm th}}\frac{q_0q_2}{q_1^2}\right)^{1/2} = \nonumber\\ & = & T_{\rm min}^{(1)}+\frac{(q_0q_2)^{1/2}}{q_1} T_{\rm min}^{(C)} . \label{Eq:tmin}
\end{eqnarray}
For perfect photon number resolving detection, when $q_0 q_2 = 0$, the minimum transmission reaches $T_{\rm min}^{(1)}$. It should be noted that if only one of the parameters $q_0$ and $q_2$ is zero, the expression (\ref{eq:chizero}) becomes singular: either $\lambda_{\rm opt}^{(H)} \rightarrow 0$ implying that the expected event rate goes to zero, or $\lambda_{\rm opt}^{(H)} \rightarrow \infty$, which means that generation of multiple pairs is not harmful if they can be perfectly sifted out. On the other hand, inserting $q_0 = q_1 = q_2 =1$ into Eq.~(\ref{Eq:tmin}) recovers in the leading order of $d_B$ the weak coherent pulse value $ T_{\rm min}^{(C)}$.

\begin{figure}[pt]
\centerline{\resizebox{1.0\textwidth}{!}{\psfig{file=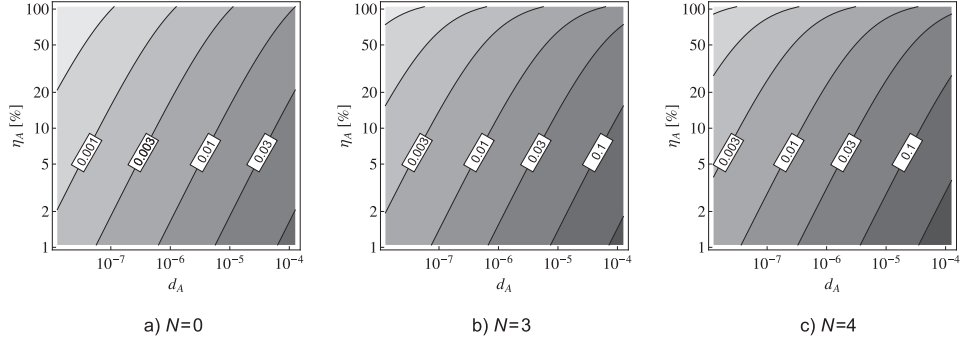}}}
\vspace*{8pt}
\caption{Contour plots of the expression $(q_0q_2)^{1/2}/ q_1$ as a function of $d_A$ and $\eta_A$ with $\eta_c=98\%$ for (a) a binary detector and multiplexed detection with (b) $N=3$ and (c) $N=4$ stages.}
\label{Fig:q0q2/q1}
\end{figure}

For realistic parameter values we have $T_{\rm min}^{(C)} \gg T_{\rm min}^{(1)}$ due to less favorable scaling with $d_B$. Therefore, the factor $(q_0 q_2 )^{1/2}/ q_1$ appearing in Eq.~(\ref{Eq:tmin}) will define the usefulness of a heralded single photon source in QKD: the lower its value, the longer the achievable secure distance. As a quantitative illustration, in Fig.~\ref{Fig:q0q2/q1}
we plot the factor $(q_0q_2)^{1/2}/q_1$ as a function of the detector efficiency $\eta_A$ and the dark count probability $d_A$ for a binary heralding detector and multiplexed detection discussed in Sec.~\ref{Sec:HighTransmission} with $N=3$ and $4$ stages.
We see in Fig.~\ref{Fig:q0q2/q1}(a) that binary heralding indeed extends substantially the secure distance (equivalently, lowers the minimum secure transmission) compared to the WCP scheme, as pointed out first by Brassard {\em et al.}\cite{BrasLutkPRL00} For multiplexed heralding shown in Figs.~\ref{Fig:q0q2/q1}(b) and \ref{Fig:q0q2/q1}(c), the factor $(q_0q_2)^{1/2}/q_1$ increases. This means that the maximum secure distance is shortened compared to binary heralding, but nevertheless remains much longer than in the WCP scheme. When $d_A\ll {\eta}_A \eta_c^N /(1-{\eta}_A \eta_c^N )$, we can approximate
\begin{equation}
\frac{(q_0q_2)^{1/2}}{q_1}  \approx d_A^{1/2} \left(1+2^{N+1}\frac{1-{\eta}_A\eta_c^N}{{\eta}_A\eta_c^N}\right)^{1/2}.
\end{equation}
When comparing this result with binary detection ($N=0$), we  see that indeed the increased rate of dark count events resulting from multiplying the number of detectors overrides the benefits of photon number resolution, with the exception of an unrealistic case when ${\eta}_A\eta_c^N = 100\%$.

Although multiplexing fails to enhance the maximum secure distance achievable with heralded sources, it can bring benefits in the intermediate regime when the WCP scheme is no longer secure. Namely, if the transmission of the channel remains below $T^{(C)}_{\rm min}$, but above the value $T_{\rm min}^{(H)}$ for multiplexed heralding, results of Sec.~\ref{Sec:HighTransmission} suggest that multiplexing can be used as a relatively straightforward way to increase the key rate above that offered by binary heralding. Numerical results presented in the next section will confirm this observation.

\begin{figure}[pt]
\centerline{\resizebox{0.8\textwidth}{!}{\psfig{file=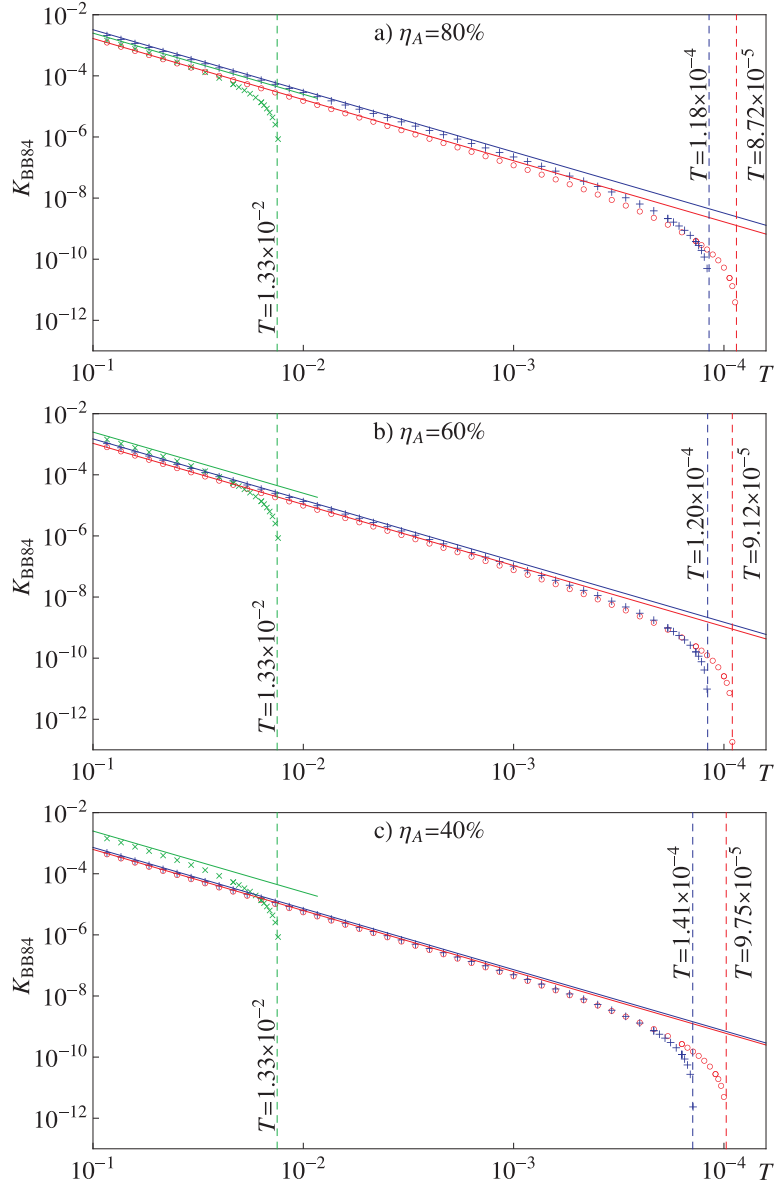}}}
\vspace*{8pt}
\caption{Log-log plots of key rates for the BB84 protocol as a function of the overall channel transmission $T$ calculated numerically for (a) $\eta_A = 80\%$, (b) $\eta_A = 60\%$, and (c) $\eta_A = 40\%$. Numerical results are shown for weak coherent pulses ($\times$, green online), a binary heralding detector ($\circ$, red online), and a multiplexed heralding detector ($+$, blue online) with the number of stages maximizing the key rate in the short distance regime ($N=4$ for (a), $N=3$ for (b) and (c)). Other parameter values are $\eta_c=98\%$, $d_A=10^{-6}$, and $d_B=10^{-5}$. Solid lines depict the analytical approximations of the key rates in the short distance regime obtained from Eq.~(\protect\ref{Eq:KHSPSapprox}). Vertical dashed lines depict minimum transmissions calculated using Eq.~(\protect\ref{Eq:tmin}). Line colors online correspond to those of the numerical results.}
\label{Fig:KBB84}
\end{figure}

\section{Numerical results}
\label{Sec:Numerical}

In order to gain a more complete picture of the performance of analyzed QKD protocols, we will now present numerical optimization of the key rate given in Eq.~(\ref{Eq:KeyRate}) over $\lambda$, which has the interpretation of the average photon number for the weak coherent pulse implementation and  the pair production probability for a heralded single photon source. This optimization needs to be carried out individually for each overall transmission $T$ of the optical channel. As a model for the photon number resolving detector we use the multiplexing scheme described in Sec.~\ref{Sec:HighTransmission}.

The results for the BB84 protocol are shown in Fig.~\ref{Fig:KBB84}. As a reference, we plot key rates attainable with WCPs and a heralded source employing a binary on/off detector. As analyzed in Sec.~\ref{Sec:HighTransmission}, the critical parameter is the efficiency $\eta_A$ of the heralding detector. For sufficiently high efficiencies, multiplexing is beneficial over a wide range of transmission, including the entire range when the weak coherent pulse implementation offers security. For lower $\eta_A$, the optimal scheme over short distances (i.e.\ high transmission) is to use WCPs. However, the corresponding security range is rather limited. Beyond that limit, for intermediate distances, multiplexing provides an advantage until security becomes compromised by increased dark count rates. In the regime of very low transmissions, a heralded source with a binary detector is the only setup that still warrants security.

\begin{figure}[pt]
\centerline{\resizebox{0.8\textwidth}{!}{\psfig{file=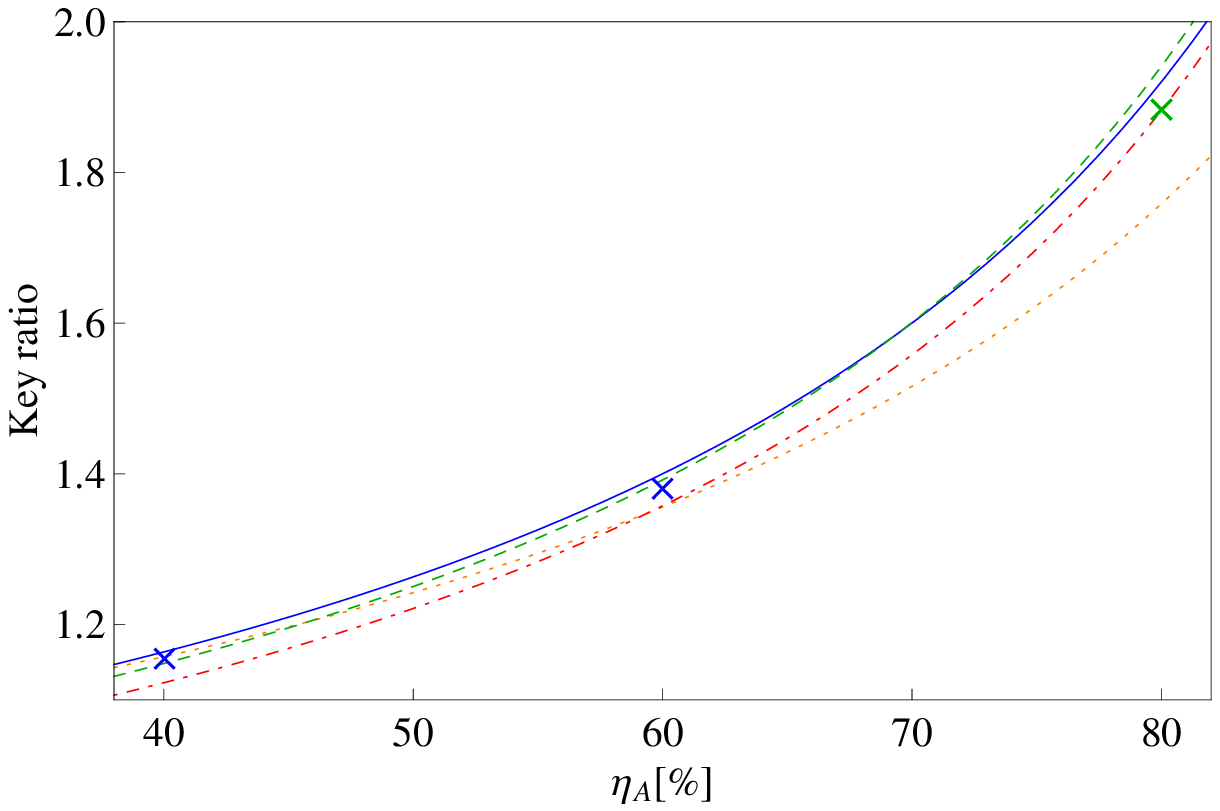}}}
\vspace*{8pt}
\caption{The ratio of key rates for an $N$-stage mutliplexed heralding detector and a binary detector as a function of $\eta_A$ calculated using Eq.~(\ref{Eq:q12q2}) for $\eta_c=98\%$, $d_A=10^{-6}$, and $N=2$ (dotted line, orange online) $N=3$ (solid line, blue online), $N=4$ (dashed line, green online) and $N=5$ (dot-dashed line, red online). Symbols $\times$ depict results obtained by fitting numerical data (using linear regression method) to the model $K\propto T^2$ for the optimal numbers of splitting stages: $N=3$ for  $\eta_A=40\%$ and $\eta_A=60\%$, and $N=4$ for $\eta_A=80\%$.}
\label{Fig:KeyRatio}
\end{figure}

The plots show that over a large part of the intermediate distance the approximation from Sec.~\ref{Sec:HighTransmission} still holds. Therefore Eq.~(\ref{Eq:q12q2}) provides a simple method to estimate the increase in the key rate when a multiplexing detector is used in lieu of a binary one and to find the optimal number of multiplexing stages when couplers are lossy. In Fig.~\ref{Fig:KeyRatio} we depict the ratio of the key rates for an $N$-stage mutliplexed heralding and binary heralding as a function of $\eta_A$ calculated using Eq.~(\ref{Eq:q12q2}), assuming non-unit transmission of couplers equal to $\eta_c=98\%$. The optimum is achieved for either $N=3$ or $N=4$ stages, depending on the efficiency $\eta_A$. Fig.~\ref{Fig:KeyRatio} also shows that this simple analytical derivation agrees quite well with enhancements determined from fitting numerical results in the short-distance regions where the key rate remains approximately proportional to $T^2$.

\begin{figure}[pt]
\centerline{\resizebox{0.8\textwidth}{!}{\psfig{file=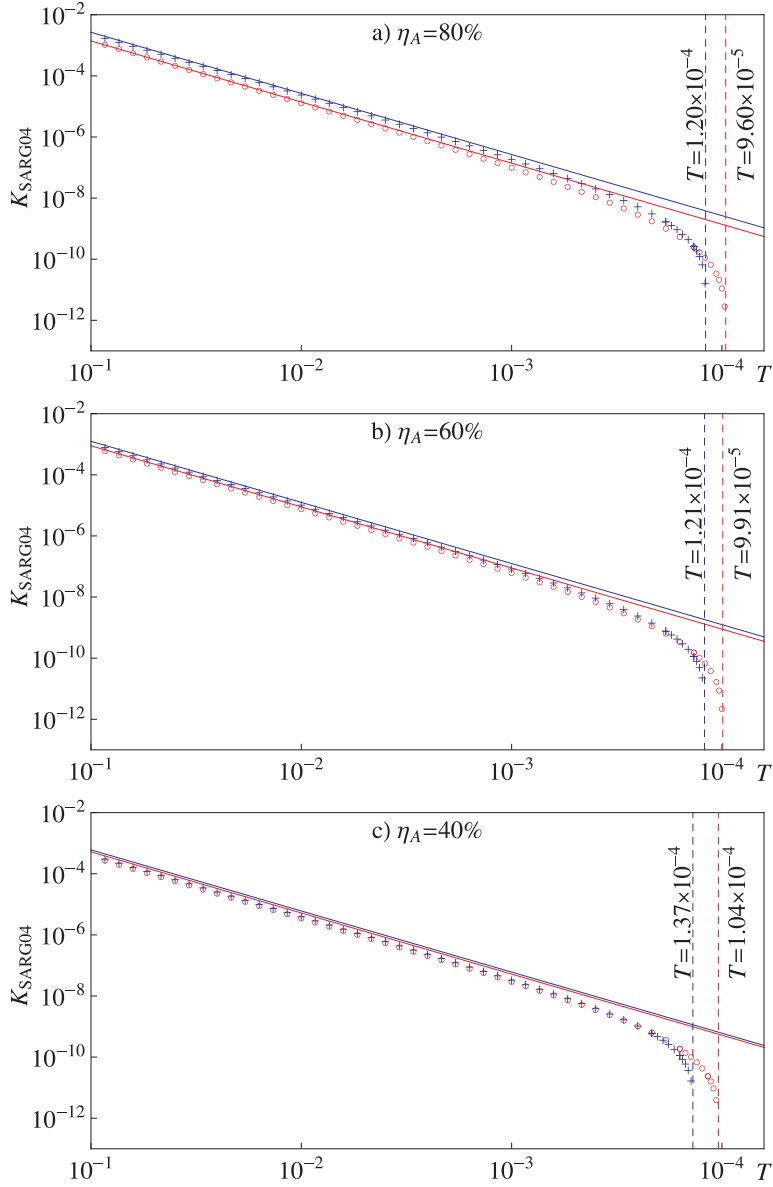}}}
\vspace*{8pt}
\caption{Log-log plots of key rates for the SARG04 protocol as a function of the overall channel transmission $T$ calculated numerically for (a) $\eta_A = 80\%$, (b) $\eta_A = 60\%$, and (c) $\eta_A = 40\%$. Numerical results are shown for a binary heralding detector ($\circ$, red online) and a multiplexed heralding detector ($+$, blue online) with the number of stages maximizing the key rate in the short distance regime ($N=4$ for (a), $N=3$ for (b) and (c)). Other parameter values are $\eta_c=98\%$, $d_A=10^{-6}$, and $d_B=10^{-5}$. Solid lines depict the analytical approximations of the key rates in the short distance regime obtained from Eq.~(\protect\ref{Eq:KHSPSapprox}). Vertical dashed lines depict minimum transmissions calculated using Eq.~(\protect\ref{Eq:tmin}). Line colors online correspond to those of the numerical results.}
\label{Fig:KSARG04}
\end{figure}

Fig.~\ref{Fig:KSARG04} shows analogous results for the SARG04 protocol assuming the specific eavesdropping strategy discussed in Sec.~\ref{Sec:KeyRate}. We found that for WCPs and low transmissions numerical optimization of the pulse amplitude took us outside the regime when $I_{AE}^{(1)} (Q/y) \le I_{AE}^{(2)}$, therefore we depict here only results for a heralded single photon source and compare performance of multiplexed and binary heralding. It is seen that again multiplexing is more beneficial for high heralding efficiencies but leads to a reduction of the maximum secure distance.

\section{Conclusions}
\label{Sec:Conclusions}

In this paper we analyzed the potential advantages of realistic heralded single photon sources in quantum key distribution based on four-state qubit protocols. We identified the characteristics of detectors used for heralding that is relevant to the performance of QKD protocols, including the key rates in the high-transmission limit when dark counts can be neglected, as well as the minimum transmission that still ensures key security. This can be used to evaluate the usefulness of new approaches to achieve photon number resolution,\cite{EisaFanRSI11,HadfNPh09} such as multipixel\cite{ThomYuanNCo12,KalaTanOE12} or superconducting\cite{JahaFrucAPL12} detectors.

The results were specialized to the case of multiplexing detectors, which offer partial photon number resolution while adding relatively little extra complexity to the experimental setup compared to binary on/off detection. We found that increasing the key rate above that attainable with weak coherent pulses poses stringent requirements on the overall efficiency in the heralding arm of the SPDC setup, including collection efficiency of down-converted photons. However, the implementation based on weak coherent pulses has a very restricted security range, and beyond this limit multiplexing can increase the key rate compared to binary heralding. Unfortunately, for very long distances (i.e. low transmissions), the increased dark count rate of the multiplexing scheme jeopardizes security leaving the single binary detector as the only option.

\section*{Acknowledgments}

We acknowledge insightful discussions with \mbox{F. Grosshans}, \mbox{N. L\"{u}tkenhaus} and \mbox{R. Thew}. This research was supported by the FP7 FET project Q-ESSENCE (Grant Agreement \mbox{no.\ 248095}), the Foundation for Polish Science TEAM project cofinanced by the EU European Regional Development Fund, and Polish NCBiR under the ERA-NET CHIST-ERA project QUASAR.


\begin{thebibliography}{30}

\bibitem{BB84}
C. H. Bennett and G. Brassard, Quantum Cryptography: Public key distribution and coin
 tossing in {\it Proc. IEEE Int. Conf. on Computers, Systems, and Signal Processing}, Bangalore, India (1984), pp.~175--179.

\bibitem{GisiRiboRMP02}
N. Gisin, G. Ribordy, W. Tittel and H. Zbinden, {\it Rev. Mod. Phys.} {\bf 74} (2002) 145--195.

\bibitem{EisaFanRSI11}
M.D. Eisaman, J. Fan, A. Migdall and S.V. Polyakov, {\it Rev. Sci. Instrum.} {\bf 82} (2011) 071101.

\bibitem{ScarBechRMP09}
V. Scarani {\it et al}., {\it Rev. Mod. Phys.} {\bf 81} (2009) 1301--1350.

\bibitem{BrasLutkPRL00}
G. Brassard, N. L\"{u}tkenhaus, T. Mor and B.C. Sanders, {\it Phys. Rev. Lett.} {\bf 85} (2000) 1330--1333.

\bibitem{BridDegiAPL12}
G. Brida {\it et al}., {\it Appl. Phys. Lett.} {\bf 101} (2012) 221112.

\bibitem{KurtOberPRA01}
C. Kurtsiefer, M. Oberparleiter and H. Weinfurter, {\it Phys. Rev. A} {\bf 64} (2001) 023802.

\bibitem{LjunTengPRA05}
D. Ljunggren and M. Tengner, {\it Phys. Rev. A} {\bf 72} (2005) 062301.

\bibitem{KoleWasiPRA09}
P. Kolenderski, W. Wasilewski and K. Banaszek, {\it Phys. Rev. A} {\bf 80} (2009) 013811.

\bibitem{SARG04}
V. Scarani, A. Ac\'{i}n, G. Ribordy and N. Gisin, {\it Phys. Rev. Lett.} {\bf 92} (2004) 057901.

\bibitem{HwangPRL03}
W.-Y. Hwang, {\it Phys. Rev. Lett.} {\bf 91} (2003) 057901.

\bibitem{WangPRL05}
X.-B. Wang, {\it Phys. Rev. Lett.} {\bf 94} (2005) 230503.

\bibitem{LoMaChenPRL05}
H.-K. Lo, X. Ma and K. Chen, {\it Phys. Rev. Lett.} {\bf 94} (2005) 230504.

\bibitem{MaQiPRA05}
X. Ma, B. Qi, Y. Zhao and H.-K. Lo, {\it Phys. Rev. A} {\bf 72} (2005) 012326.

\bibitem{MaurSilbPRA07}
W. Mauerer and C. Silberhorn, {\it Phys. Rev. A} {\bf 75} (2007) 050305.

\bibitem{AdaYamaPRL07}
Y. Adachi, T. Yamamoto, M. Koashi and N. Imoto, {\it Phys. Rev. Lett.} {\bf 99} (2007) 180503.

\bibitem{MaLoNJP08}
X. Ma and H.-K. Lo, {\it New J. Phys.} {\bf 10} (2008) 073018.

\bibitem{CurtyMaPRA10}
M. Curty, X. Ma, B. Qi and T. Moroder, {\it Phys. Rev. A} {\bf 81} (2010) 022310.

\bibitem{HoriKobaPRA06}
T. Horikiri and T. Kobayashi, {\it Phys. Rev. A} {\bf 73} (2006) 032331.

\bibitem{WangChenPRL08}
Q. Wang {\it et al}., {\it Phys. Rev. Lett.} {\bf 100} (2008) 090501.

\bibitem{HelwMauePRA09}
W. Helwig, W. Mauerer and C. Silberhorn, {\it Phys. Rev. A} {\bf 80} (2009) 052326.

\bibitem{TanCaiIJQI11}
Y.-G. Tan and Q.-Y. Cai, {\it Int. J. Quant. Inf.} {\bf 9} (2011) 903.

\bibitem{FiniteKeyLength}
V. Scarani and R. Renner, {\it Phys. Rev. Lett.} {\bf 100} (2008) 200501.

\bibitem{RennerIJQI08}
R. Renner, Int. J. Quant. Inf. {\bf 6} (2008) 1.

\bibitem{GLLP}
D. Gottesman, H.-K. Lo, N. L\"{u}tkenhaus and J. Preskill {\it Quantum Inf. Comput.} {\bf 4} (2004) 325--360.

\bibitem{BranGisiPRA05}
C. Branciard, N. Gisin, B. Kraus and V. Scarani, {\it Phys. Rev. A} {\bf 72} (2005) 032301.

\bibitem{Holevo}
A.S. Holevo, {\it Probl. Inf. Transm.} {\bf 9} (1973) 177--183.

\bibitem{TamaLoPRA06}
K. Tamaki and H.-K. Lo, {\it Phys. Rev. A} {\bf 73} (2006) 010302.

\bibitem{AchiSilbOL03}
D. Achilles, C. Silberhorn, C. \'{S}liwa, K. Banaszek and I.A. Walmsley, {\it Opt. Lett.} {\bf 28} (2003) 2387--2389.

\bibitem{FitcJacoPRA03}
M.J. Fitch, B.C. Jacobs, T.B. Pittman and J.D. Franson, {\it Phys. Rev. A} {\bf 68} (2003) 043814.

\bibitem{HadfNPh09}
R.H. Hadfield, {\it Nature Photon.} {\bf 3} (2009) 696--705.

\bibitem{ThomYuanNCo12}
O. Thomas, Z.L. Yuan and A.J. Shields, {\it Nat. Commun.} {\bf 3} (2012) 644.

\bibitem{KalaTanOE12}
D.A. Kalashnikov, S.-H. Tan and L.A. Krivitsky, {\it Opt. Express} {\bf 20} (2012) 5044--5051.

\bibitem{JahaFrucAPL12}
S. Jahanmirinejad {\it et al}., {\it Appl. Phys. Lett.} {\bf 101} (2012) 072602.

\end{thebibliography}
\end{document}